\documentclass[twocolumn,showpacs,preprintnumbers,amsmath,amssymb]{revtex4}
\usepackage{graphicx}% Include figure files
\usepackage{dcolumn}% Align table columns on decimal point
\usepackage{bm}% bold math

\begin{document}

%\preprint{APS/123-QED}

\title{
Low-temperature x-ray crystal structure analysis of the cage-structured compounds $M$Be$_{13}$ ($M$ = La, Sm, and U)
}% Force line breaks with \\

\author{Hiroyuki Hidaka} 

\email{hidaka@phys.sci.hokudai.ac.jp}, 

\author{Ryoma Nagata} 

\author{Chihiro Tabata} 

\thanks{Present address: Condensed Matter Research Center and Photon Factory, Institute of Materials Structure Science, High Energy Accelerator Research Organization, Tsukuba, Ibaraki 305-0801, Japan}

\author{Yusei Shimizu} 

\thanks{Present address: Institute for Materials Research, Tohoku University, Oarai, Ibaraki 311-1313, Japan}

\author{Naoyuki Miura} 

\author{Tatsuya Yanagisawa} 

\author{Hiroshi Amitsuka} 

\affiliation{Graduate School of Science, Hokkaido University, Sapporo, Hokkaido 060-0810, Japan}%

\date{\today}% It is always \today, today,
% but any date may be explicitly specified

\begin{abstract}
The beryllides $M$Be$_{13}$ ($M$ = rare earths and actinides) crystallize in a NaZn$_{13}$-type cubic structure, which can be categorized as a cage-structured compound. 
In this study, powder x-ray diffraction measurements have been performed on LaBe$_{13}$, SmBe$_{13}$, and UBe$_{13}$ in the temperature range between 7 and 300 K in order to investigate their crystallographic characteristics systematically. 
They keep the NaZn$_{13}$-type cubic structure down to the lowest temperature. 
We estimated their Debye temperature to be 600--750 K from analyses of the temperature dependence of a lattice parameter, being in good agreement with the values reported previously. 
Rietveld refinements on the obtained powder patterns revealed that the $M$ atom in the 8$a$ site is located in an almost ideal snub cube formed by 24 Be$^{\rm II}$ atoms in the 96$i$ site, whose caged structure is unchanged even at the low temperatures. 
In addition, it is argued from the temperature variation of an isotropic mean-square displacement parameter that the $M$Be$_{13}$ compounds commonly have a low-energy phonon mode, which can be described by a model assuming an Einstein oscillation of the $M$ atom with a characteristic temperature of $\sim$ 160 K. 
\end{abstract}

%\pacs{71.27.+a, 75.20.Hr, 71.30.+h, 74.62.Fj}

\maketitle

\section{INTRODUCTION}

For the past several decades, intermetallic compounds with a caged structure have been intensively studied as potential candidates for thermoelectric conversion applications. \cite{Sales1, Suekuni, Sales2} 
In the research field of the strongly correlated electron systems, cage-structured compounds, e.g., filled skutterudites, clathrates, $\beta$-pyroclores, hexaborides, and Al$_{10}$V, have attracted much attention because of their novel phenomena, such as higher-rank multipole ordering in PrRu$_4$P$_{12}$ \cite{Takimoto}, a magnetic-field-robust heavy-fermion (HF) state in SmOs$_4$Sb$_{12}$ \cite{Sanada}, and a low-energy phonon mode associated with local vibration of a guest atom with a large amplitude in an oversized host cage, so-called “rattling”. \cite{Suekuni, Matsuhira, Yamaura1, Yamaura2, Mandrus, Caplin}
Among them, the rattling has been considered to be related with several intriguing phenomena via interaction with conduction electrons, for example, a rattling-induced superconductivity \cite{Nagao}, unusual temperature dependence of the electrical resistivity \cite{Dahm}, and a magnetic-field-insensitive HF state due to the multi-level Kondo effect. \cite{Hattori} 
In contrast to the previous expectations, it has recently been claimed from neutron spectroscopy experiments that a picture of a freely “rattling” guest atom in a host cage is not applicable in the case of filled Fe$_4$Sb$_{12}$ skutterudites. \cite{Koza} 
Thus, the issue of rattling is still controversial. 
To verify the underlying concept of rattling itself, it is highly necessary to examine crystallographic characteristics at low temperatures for several cage-structured systems.

\begin{figure}[htb]
\begin{center}
\includegraphics[width=0.65\linewidth]{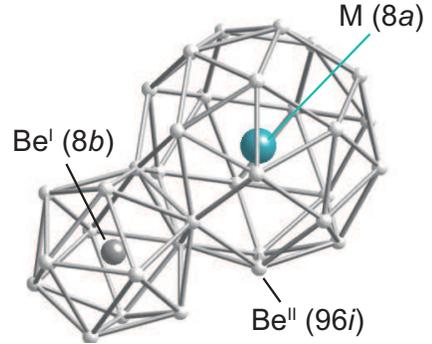}
\caption[]{\protect (Color online) Cagelike structures of $M$Be$_{13}$. The $M$ and Be$^{\rm I}$ atoms are placed into an almost ideal snub cube and an icosahedron formed by the Be$^{\rm II}$ atoms, respectively. } 
\end{center}
\end{figure}

The beryllide $M$Be$_{13}$ can also be categorized as a cage-structured compound, where $M$ is a rare earth or an actinide. 
They crystallize in a NaZn$_{13}$-type cubic structure with the space group $F$$m$$\bar{\rm 3}$$c$ (No. 226, $O_h^{\rm 6}$), where the unit cell contains $M$ atoms in the 8$a$ site, Be$^{\rm I}$ atoms on the 8$b$ site, and Be$^{\rm II}$ atoms in the 96$i$ site. \cite{Bucher, McElfresh, Takegahara} 
The $M$ atoms are surrounded by 24 Be$^{\rm II}$ atoms, nearly forming an ideal snub cube, whereas the Be$^{\rm I}$ atoms are surrounded by 12 Be$^{\rm II}$ atoms, forming an icosahedron cage, as shown in Fig. 1.

For the phonon physics in $M$Be$_{13}$, previous specific-heat and inelastic neutron scattering (INS) measurements revealed that UBe$_{13}$, ThBe$_{13}$, and possibly LaBe$_{13}$ have a low-energy phonon mode, which can be described by an Einstein mode with characteristic temperatures of $\theta_{\rm E}$ = 151, 157, and 177 K, respectively. \cite{Renker, Hidaka} 
To discuss a possibility of an influence of the low-energy phonon mode on novel phenomena in the $M$Be$_{13}$ system, such as unconventional HF superconductivity and non-Fermi-liquid behavior in UBe$_{13}$ \cite{Ott, Mayer} and an intermediate valence state due to the strong $c$-$f$ hybridization in CeBe$_{13}$ \cite{Wilson}, it is necessary to elucidate characteristics of the low-energy phonon mode. 
It has been suggested that the $M$ atom behaves like an Einstein oscillator in UBe$_{13}$ and ThBe$_{13}$ \cite{Renker}, while other characteristics of the low-energy phonon mode in the present system have been veiled: for example, coupling between the guest ions and the host lattice, form of the potential energy (harmonic or anharmonic), and size effects of the cage. 
Systematic crystallographic studies on the low-energy phonon mode for the $M$Be$_{13}$ series would provide important information to unveil the universal nature of the phonon properties in $M$Be$_{13}$ with unique cagelike structures.

A low-temperature systematic investigation of crystal structure is also crucial to unveil the physical ground-state properties, such as a gap symmetry of the unconventional superconductivity in UBe$_{13} $\cite{Sigrist} and crystalline-electric-field effects in all the magnetic $M$Be$_{13}$ compounds, both of which may be affected by some crystal distortion.　A lattice parameter $a$ at room temperature (RT) has already been reported in all the $M$Be$_{13}$ compounds \cite{Bucher, Benedict}, while the low-temperature data are limited in some $M$Be$_{13}$ compounds \cite{Kappler, Goldman}. 
The fractional (0, $y$, $z$) coordinates of the Be$^{\rm II}$ site at RT have been determined in several $M$Be$_{13}$ ($M$ = La, Ce, Dy, Th, and U) \cite{Hidaka, Goldman, Hudson}, whereas their low-temperature values have been determined only in UBe$_{13}$ and ThBe$_{13}$. \cite{Goldman} 
In addition, studies on an atomic displacement parameter for $M$Be$_{13}$, which can be a useful guide to understand phonon properties \cite{Suekuni, Yamaura1}, have remained untouched even at RT except for UBe$_{13}$ and ThBe$_{13}$ \cite{Goldman}.

In this study, we investigated structure parameters and phonon properties of a superconductor LaBe$_{13}$ with a superconducting transition temperature of $\sim$0.6 K \cite{Hidaka}, a possible helical magnet SmBe$_{13}$ with a magnetic transition temperature of $T_{\rm M}$ = 8.3 K \cite{Bucher, Hidaka2, Mombetsu}, and an unconventional HF superconductor UBe$_{13}$ \cite{Ott} by x-ray diffraction (XRD) measurements in temperatures of down to 7 K. 
The results of the XRD measurements and Rietveld analyses are presented in Sec. III: the temperature dependence of the lattice parameter in subsection A, the fractional (0, $y$, $z$) coordinates in subsection B, and isotropic mean-square displacement parameters in subsection C.

\section{EXPERIMENT}

Single crystals of LaBe$_{13}$, SmBe$_{13}$, and UBe$_{13}$ were grown by an Al-flux method. 
The constitute materials were La with 99.9$\%$ purity, Sm with 99.9$\%$ purity, U with 99.5$\%$ purity, Be with 99$\%$ purity, and Al with 99.99$\%$ purity. 
Each M element was placed in an Al$_2$O$_3$ crucible with Be and Al at an atomic ratio of La:Be:Al = 1:13:35, Sm:Be:Al = 1:13:30, and U:Be:Al = 1:13:40, and sealed in a quartz tube filled with ultrahigh-purity Ar gas of 150 mmHg. 
The sealed tube was kept at 1050 $^\circ$C for 1 week and then cooled at a rate of 2 $^\circ$C/h. 
The Al flux was spun off in a centrifuge and then removed by NaOH solution. 
The obtained single crystals were annealed for 2 weeks at 700 $^\circ$C. 
The annealed samples were ground into fine powders in the ethyl alcohol, and then put in a cupper holder for XRD.

The XRD measurements were performed by a conventional powder x-ray diffractometer (Rint 2000, Rigaku) using Cu K$\alpha_{1}$ and K$\alpha_{2}$ radiation in the angle range of $15^\circ$ $<$ 2$\theta$ $<$ $157^\circ$. 
A Gifford-McMahon refrigerator was used for the low-temperature measurements down to 7 K. 
Rietveld refinement on the powder XRD data was carried out using the RIETAN-FP program. \cite{Rietan}

\begin{figure}[htb]
\begin{center}
\includegraphics[width=1.0\linewidth]{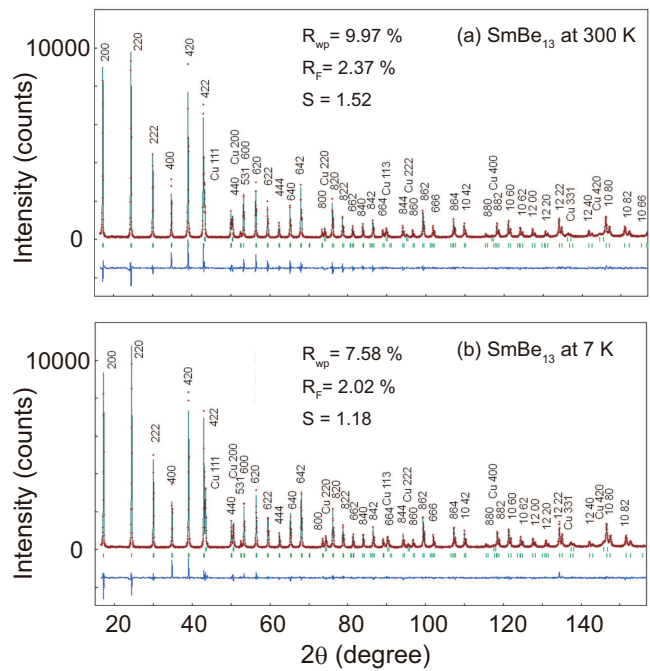}
\caption[]{\protect (Color online) Observed powder XRD patterns of SmBe$_{13}$ (cross symbols) and calculated profiles (solid lines) at (a) 300 and (b) 7 K. 
The green vertical bars below the XRD patterns indicate the calculated peak positions. 
The difference between the observed and calculated intensities is indicated by the blue line at the bottom of each figure.} 
\end{center}
\end{figure}

\section{RESULT AND DISCUSSION}

\subsection{Lattice parameters}

\begin{figure}[htb]
\begin{center}
\includegraphics[width=0.8\linewidth]{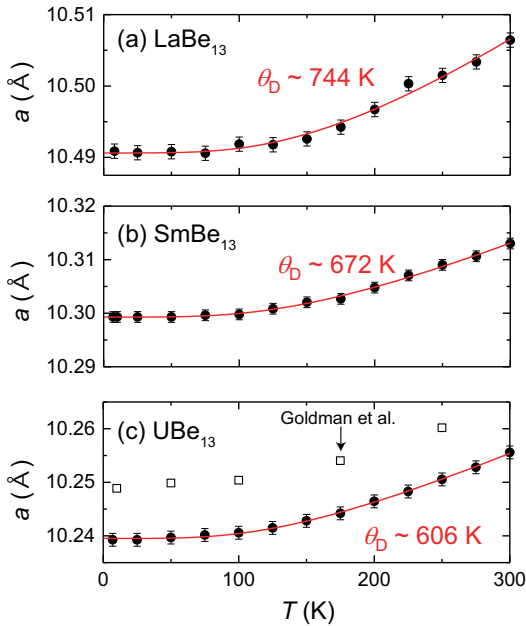}
\caption[]{\protect (Color online) Temperature dependence of the lattice parameter $a$($T$) for (a) LaBe$_{13}$, (b) SmBe$_{13}$, and (c) UBe$_{13}$. The solid lines represent the fitting curves based on the Debye-Gr$\ddot{\rm u}$neisen approximation, as described in the text. 
The open symbols in Fig. 3(c) are the data reported previously \cite{Goldman}. } 
\end{center}
\end{figure}

\begin{table*}[htb]
\begin{center}
\caption[]{\protect Lattice parameter $a$, fractional (0, $y$, $z$) coordinates of the Be$^{\rm II}$ atom, typical reliability factors, i.e., $R_{\rm wp}$, $R_{\rm F}$, and $S$, in the Rietveld analysis for LaBe$_{13}$, SmBe$_{13}$, and UBe$_{13}$ at 300 K and the lowest temperature (8 K for LaBe$_{13}$, and 7 K for SmBe$_{13}$ and UBe$_{13}$).
The previously reported data are also listed for comparison in this table. \cite{Bucher, McElfresh, Goldman}} 
\includegraphics[width=1.0\linewidth]{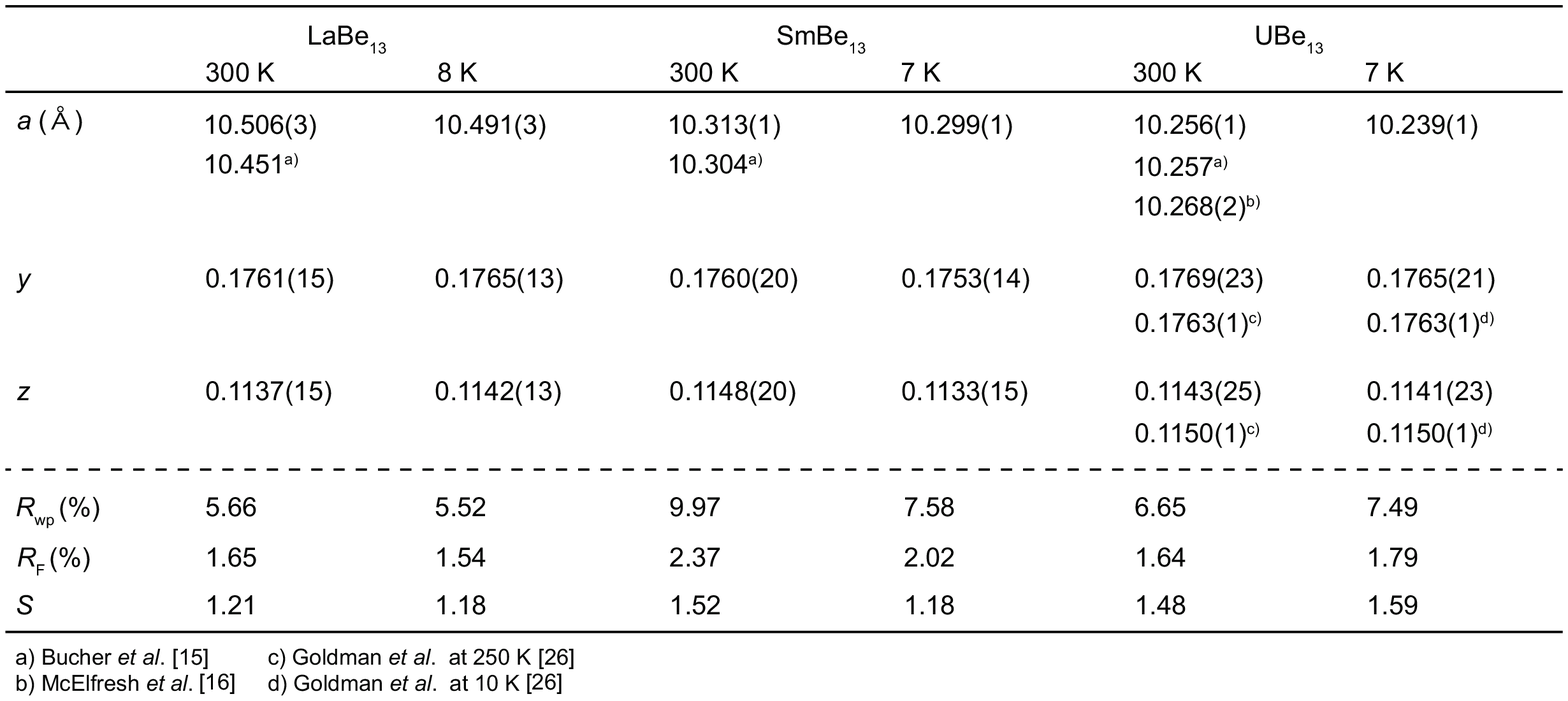}
\end{center}
\end{table*}

Figure 2 shows the powder XRD patterns of SmBe$_{13}$ at (a) 300 and (b) 7 K. 
The obtained XRD patterns can be well fitted assuming the NaZn$_{13}$-type cubic structure. 
In addition, we do not observe any peaks associated with impurity phases except for reflections from the cupper holder. 
We note that any residual peaks are smaller than 0.3$\%$ of the principal reflection. 
In the present XRD measurements, we succeeded in observing weak signals, which originate from only the Be$^{\rm II}$ atoms, for example (531) reflection, promising determination of the fractional coordinates of the Be$^{\rm II}$ atom. 
In addition, there is no obvious difference in the XRD patterns between 300 and 7 K except for a shift of the peak positions on the 2$\theta$ axis due to a change in the lattice parameter. 
Similar results were obtained in LaBe$_{13}$ and UBe$_{13}$ (not shown), indicating absence of a structural transition in this temperature region. 
The lattice parameter $a$ was determined from the obtained XRD peaks at high 2$\theta$ angles (2$\theta$ $>$ $110^\circ$) using the Bragg's law. 
The obtained lattice parameter at 300 K and at the lowest temperature for LaBe$_{13}$, SmBe$_{13}$, and UBe$_{13}$ are shown in Table I, which are approximately consistent with those reported previously. \cite{Bucher, McElfresh}

Figure 3 shows the temperature dependence of the lattice parameter $a$($T$) for (a) LaBe$_{13}$, (b) SmBe$_{13}$, and (c) UBe$_{13}$. 
For all three compounds investigated here, the lattice parameter decreases as the temperature is lowered, and levels off below $\sim$100 K. 
The data of UBe$_{13}$ reported previously are also indicated as the open symbols in Fig. 3(c) \cite{Goldman}. 
Although the lattice parameter of our UBe$_{13}$ sample is slightly smaller than that reported in Ref.[18], the observed temperature variation is in good agreement with the previous result. 
Note that there is no anomaly in the $a$($T$) curve near $T_{\rm M}$ of SmBe$_{13}$ within the experimental accuracy. 
From almost linear temperature dependence in the $a$($T$) curves above $\sim$150 K, we estimated a thermal expansion coefficient $\alpha$ ($\equiv$ 1/$a$ d$a$/d$T$). 
The estimated $\alpha$ values at 300 K are 8.8 $\times$ 10$^{-6}$, 7.1 $\times$ 10$^{-6}$, and 8.3 $\times$ 10$^{-6}$ K$^{-1}$ for LaBe$_{13}$, SmBe$_{13}$, and UBe$_{13}$, respectively, as summarized in Table II. 
Similar $a$($T$) has also been reported in YBe$_{13}$, PrBe$_{13}$, LuBe$_{13}$, and ThBe$_{13}$, where $\alpha$ can be estimated to be 7 -- 9 $\times$ 10$^{-6}$ K$^{-1}$. \cite{Kappler} 
It is revealed that $\alpha$ above $\sim$ 150 K in the $M$Be$_{13}$ system is almost independent of the $M$ atom.

To estimate a Debye temperature $\theta_{\rm D}$, the $a$($T$) curves were analyzed using the Debye-Gr$\ddot{\rm u}$neisen approximation \cite{Sayetat} described by 
\begin{equation}
a(\it T) \cong a_{\rm 0} \Biggl[\rm 1 + 3\it I_a T \Biggl( \frac{T}{\theta_{\rm D}} \Biggr)^{\rm 3} \int_{\rm 0}^{\theta_{\rm D}/T} \frac{x^{\rm 3}}{\rm exp(\it x) - \rm1} dx \Biggr], 
\end{equation}
\begin{equation}
I_a = \frac{k_{\rm B}G}{BV_{\rm 0}}, 
\end{equation}
where $k_{\rm B}$ is the Boltzmann constant, $G$ is the Gr$\ddot{\rm u}$neisen parameter, $B$ is the bulk modulus, and $a_0$ and $V_0$ are a lattice parameter and volume of unit cell at absolute zero, respectively. 
The obtained fitting parameters are ($\theta_{\rm D}$, $a_0$, $I_a$) = (744 K, 10.491 $\rm \AA$, 1.4 $\times$ 10$^{-5}$) for LaBe$_{13}$, (672 K, 10.299 $\rm \AA$, 1.1 $\times$ 10$^{-5}$) for SmBe$_{13}$, and (606 K, 10.240 $\rm \AA$, 1.2 $\times$ 10$^{-5}$) for UBe$_{13}$. 
The obtained $\theta_{\rm D}$ values are listed in Table II along with the previously reported values of $\theta_{\rm D}$ for LaBe$_{13}$ and ThBe$_{13}$. \cite{Bucher, Kappler, Hidaka} 
These results indicate that the $M$Be$_{13}$ systems commonly have relatively high $\theta_{\rm D}$ of 600--900 K, although it is necessary to consider an influence of the low-energy phonon mode for more proper discussion.

\begin{table*}[t]
\begin{center}
\caption[]{\protect Thermal expansion coefficient $\alpha$ at 300 K, Debye temperature $\theta_{\rm D}$, Einstein temperature $\theta_{\rm E}$, and electronic specific-heat coefficient $\gamma$ for LaBe$_{13}$, SmBe$_{13}$, UBe$_{13}$ and ThBe$_{13}$. 
$\theta_{\rm D}$ and $\theta_{\rm E}$ of the $M$ atom were estimated from the temperature dependence of $U_{\rm eq}$. 
The previously reported data are also listed for comparison in this table \cite{Bucher, Hidaka, Kappler, Renker, Ott}. } 
\includegraphics[width=0.8\linewidth]{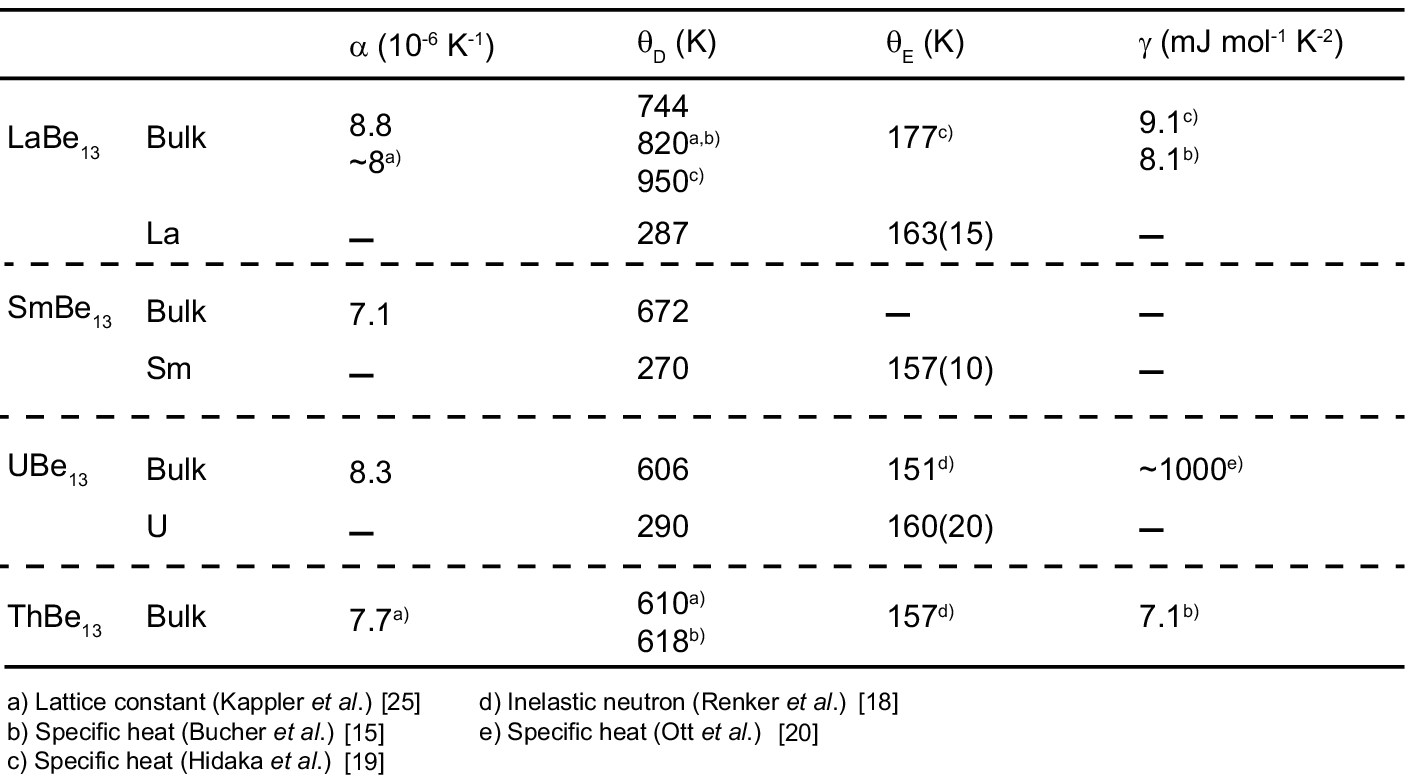} 
\end{center}
\end{table*}

\subsection{Fractional (0, $y$, $z$) coordinates of the Be$^{\rm II}$ site}

\begin{figure}[htb]
\begin{center}
\includegraphics[width=0.9\linewidth]{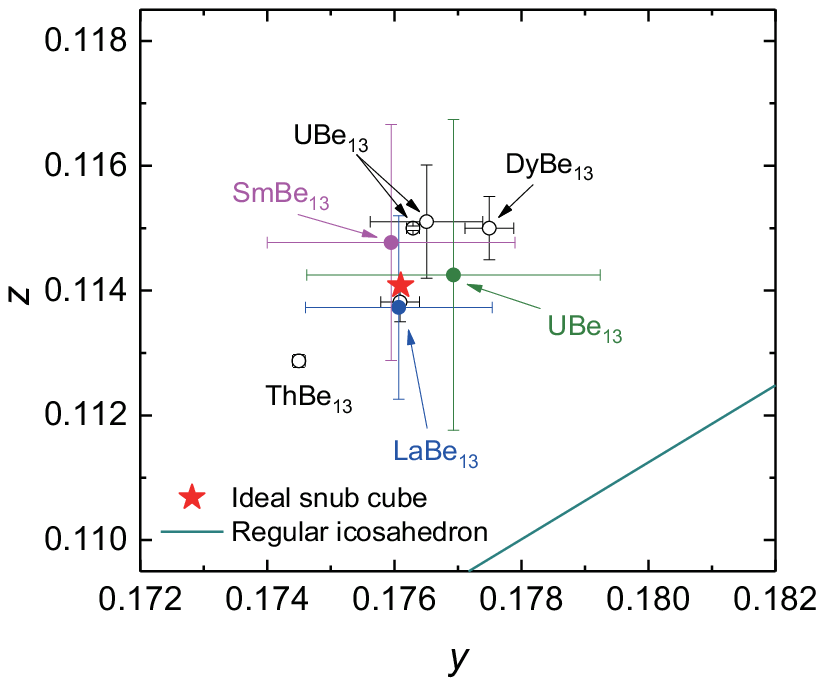}
\caption[]{\protect (Color online) Parameter plot of $y$ and $z$ at RT for several $M$Be$_{13}$ compounds ($M$ = La, Sm, Dy, Th, and U). 
The closed and open circle symbols indicate the data obtained from the present and previous studies \cite{Goldman, Hudson}, respectively. 
The fractional coordinates of the Be$^{\rm II}$ atom for $M$Be$_{13}$ are much closer to the ideal snub-cube coordinates (star symbol) than the icosahedron coordinates (solid line). 
} 
\end{center}
\end{figure}

We next show the Rietveld analyses of the obtained XRD patterns assuming the NaZn$_{13}$-type cubic structure. 
The calculated profiles for SmBe$_{13}$ can be seen in Figs. 2(a) and 2(b). 
Here, the lattice parameter was fixed to the value determined by the analysis of the Bragg's peak scan for the higher angle spectrum. 
The obtained $y$ and $z$ parameters in the fractional coordinates of the Be$^{\rm II}$ site and the typical reliability factors ($R_{\rm wp}$, $R_{\rm F}$, and $S$) are shown in Table I. 
Figure 4 shows the $y$ and $z$ parameters of the Be$^{\rm II}$ site at R.T. for several $M$Be$_{13}$ compounds: $M$ = La, Sm, Dy, Th, and U. \cite{Goldman, Hudson} 
The star symbol and the solid line indicate the ideal snub-cube ($y$ = 0.17610, $z$ = 0.11408) and regular icosahedron [$y$ = $\frac{1}{2}$(1+$\sqrt{5}$)$z$] coordinates, respectively.
In all the measured $M$Be$_{13}$ compounds, the $y$ and $z$ parameters are not close to those in the regular icosahedron coordinates but close to those in the ideal snub-cube coordinates. 
Furthermore, it is revealed that the $y$ and $z$ parameters are almost temperature-independent within the accuracy of the measurements in the present compounds, as shown in Table I, and hence their caged structures keep the almost ideal snub cube even at the low temperatures. 
This result indicates that the symmetry of a crystalline electric field at the $M$ site remains unchanged down to 7 K, although the cage size becomes smaller accompanied by the decrease of the lattice parameter upon cooling.

\subsection{Isotropic mean-square displacement parameter}

\begin{figure}[htb]
\begin{center}
\includegraphics[width=0.8\linewidth]{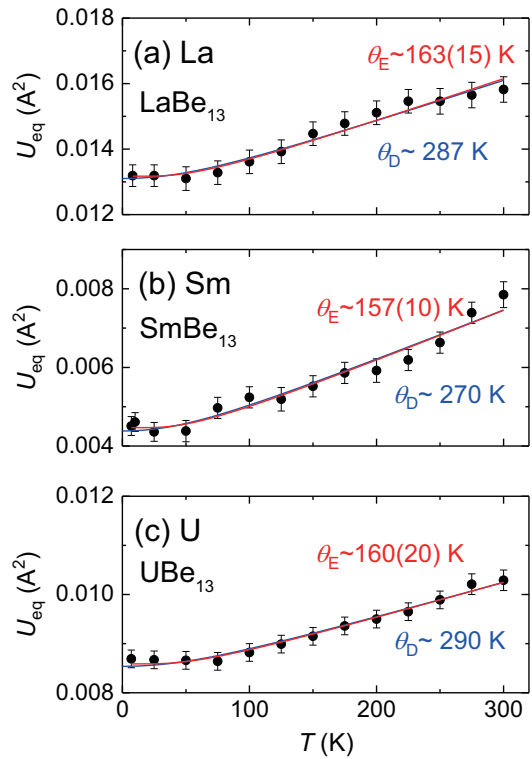}
\caption[]{\protect (Color online) Temperature dependence of $U_{\rm eq}$ of the $M$ atom for LaBe$_{13}$, SmBe$_{13}$, and UBe$_{13}$: (a) $M$ = La, (b) Sm, and (c) U. 
The red-solid and blue-solid lines indicate the fitting curves obtained from the calculations based on the Einstein and Debye models, respectively. 
}
\end{center}
\end{figure}

\begin{figure}[htb]
\begin{center}
\includegraphics[width=0.85\linewidth]{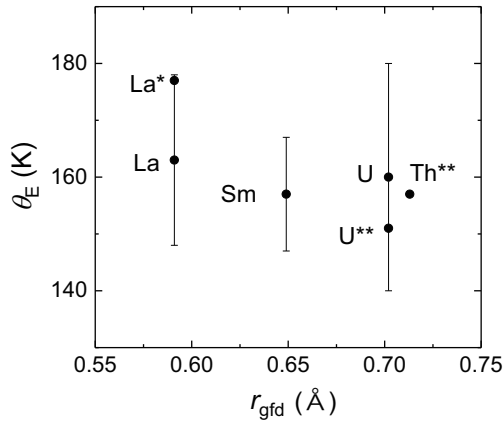}
\caption[]{\protect (Color online) Einstein temperature $\theta_{\rm E}$ vs. guest free distance $r_{\rm gfd}$ ($\equiv$ $\it{r}\rm_{M - Be^{II}}$ -- $\it{r}\rm_{M}$ -- $\it{r}\rm_{Be^{II}}$) in several $M$Be$_{13}$ compounds, where $M$ = La, Sm, Th, and U. 
Some data of $\theta_{\rm E}$, represented as La*, U** and Th**, are taken from literatures. \cite{Hidaka, Renker} 
}
\end{center}
\end{figure}

The isotropic mean-square displacement parameter $U_{\rm eq}$ of individual atoms for LaBe$_{13}$, SmBe$_{13}$, and UBe$_{13}$ were also determined from the Rietveld analyses. 
Here, an isotropic atomic displacement parameter is defined as 8$\pi^2$$U_{\rm eq}$. 
In general, it is difficult to determine an absolute value of $U_{\rm eq}$ precisely, since the precision of $U_{\rm eq}$ is strongly influenced by experimental conditions, experimental error, and refinement methods. 
In the present analyses, for example, the refined $U_{\rm eq}$ of the Sm atom for SmBe$_{13}$ took on a wide range of value from 0.003 to 0.029 depending on initial parameters at R.T. 
However, once one fixes the initial parameters to some values, the error of the refined $U_{\rm eq}$ becomes approximately $\pm$5$\%$. 
Therefore, in this study, the temperature dependence of $U_{\rm eq}$ [= $U_{\rm eq}$($T$)] was obtained by the following procedure; (1) refine $U_{\rm eq}$ at 300 K using the Rietveld analysis with some initial parameters, (2) refine $U_{\rm eq}$ at 275 K using the initial parameters determined at 300 K, (3) refine $U_{\rm eq}$ at 250 K using the initial parameters determined at 275 K, and (4) repeat the process sequentially down to the lowest temperature. 
Note that we checked that the temperature variation of $U_{\rm eq}$ is hardly influenced by the absolute value of $U_{\rm eq}$, i.e., the relative change in $U_{\rm eq}$($T$) can be used for analyses to obtain the values of $\theta_{\rm D}$ and $\theta_{\rm E}$, as discussed in the next paragraph.

The obtained $U_{\rm eq}$($T$) of the $M$ atom are shown in Fig. 5. 
Each figure represents (a) La of LaBe$_{13}$, (b) Sm of SmBe$_{13}$, and (c) U of UBe$_{13}$. 
These experimental results were analyzed using the following two models \cite{Suekuni, Bentien}: a Debye model 
\begin{equation}
U\rm_{eq}^{Deb}(\it T) = \frac{\rm 3\hbar^{\rm 2}T}{mk_{\rm B}\theta_{\rm D}^2} \Biggl[ \frac{T}{\theta_{\rm D}} \int_{\rm 0}^{\theta_{\rm D}/T} \frac{x}{\rm exp(\it x) - \rm1} dx + \frac{\theta_{\rm D}}{\rm 4 \it T} \Biggr] + \it d^{\rm 2} , 
\end{equation}
and an Einstein model
\begin{equation} 
U\rm_{eq}^{Ein}(\it T) = \frac{\hbar^{\rm 2}}{{\rm 2}mk_{\rm B}\theta_{\rm E}}\rm coth\Bigl(\frac{\theta_{\rm E}}{\rm 2\it T}\Bigr) + \it d^{\rm 2}, 
\end{equation}
where, $\hbar$, $m$ and $d^2$ are the Planck constant, mass of the atom, and temperature independent disorder term, respectively. 
The obtained $\theta_{\rm D}$ and $\theta_{\rm E}$ are listed in Table II, including the results reported previously. \cite{Hidaka, Renker} 
$U_{\rm eq}$($T$) of the $M$ atom decreases monotonously with decreasing temperature, as shown in Fig. 5. 
In the conventional Einstein model, the best fit provides $\theta_{\rm E}$ of 163(15), 157(10), and 160(20) K for the La, Sm, and U atoms, respectively, which are approximately identical to the values estimated from the specific heat ($C$) and INS measurements. \cite{Hidaka, Renker} 
On the other hand, in the Debye model, $\theta_{\rm D}$ were estimated to be $\sim$ 287, $\sim$ 270 and $\sim$ 290 K for the La, Sm, and U atoms, respectively. 
These values are less than half compared with those estimated from the $a$($T$) and $C$($T$) data. \cite{Bucher, Hidaka, Kappler} 
These results indicate that the Einstein model gives a better description of $U_{\rm eq}$($T$) of the $M$ atom than the Debye model.

In $U_{\rm eq}$($T$) of the Be$^{\rm I}$ and Be$^{\rm II}$ atoms, it was quite difficult to perform reliable analyses in both the models due to almost temperature-independent behavior of $U_{\rm eq}$ in the investigated temperature range. 
It is noted that simulated curves based on the Einstein model with $\theta_{\rm E}$ of 150 and 170 K obviously deviate from $U_{\rm eq}$($T$) of the Be$^{\rm I}$ and Be$^{\rm II}$ atoms (not shown). 
These findings suggest that the $M$ atom behaves like an Einstein oscillator with $\theta_{\rm E}$ $\sim$ 160 K, as pointed out from the previous INS measurements for UBe$_{13}$ and ThBe$_{13}$ \cite{Renker}, whereas the Be atoms form the crystal lattice described by the Debye model.

The low-energy phonon mode, which can be explained by the Einstein phonon with harmonic oscillation, is considered to be a common feature in the $M$Be$_{13}$ systems. 
It is intriguing that the $M$Be$_{13}$ series appears to have almost the same $\theta_{\rm E}$. 
In other cage-structured compounds, such as filled skutterudites \cite{Yamaura1, Matsuhira}, $\beta$-pyrochlore oxides \cite{Yamaura2}, and clathrates \cite{Suekuni}, it has been found that $\theta_{\rm E}$ shows a decreasing trend with increasing the free space for guest vibration. 
This decreasing trend in $\theta_{\rm E}$ has been interpreted as follows; larger free space yields shallower guest-ion potential, resulting in a decrease in the energy of the local phonon mode \cite{Yamaura1}. 
Here, we test the similar analysis concerning the free space, which has been performed for other cage-structured compounds. 
Figure 6 shows $\theta_{\rm E}$ for LaBe$_{13}$, SmBe$_{13}$, ThBe$_{13}$, and UBe$_{13}$, plotted against the guest free distance $r_{\rm gfd}$ determined at 300 K. \cite{Hidaka, Renker} 
In this study, $r_{\rm gfd}$ is defined as 
\begin{equation}
r\rm_{gfd} = \it{r}\rm_{M - Be^{II}} - \it{r}\rm_{M} - \it{r}\rm_{Be^{II}}, 
\end{equation} 
where, $r_{\rm M - Be^{II}}$ is the distance between $M$ and Be$^{\rm II}$, $r_{\rm M}$ is the effective ionic radius of $M$ ($r_{\rm {La^{3+}}}$ = 1.36 $\rm \AA$, $r_{\rm {Sm^{3+}}}$ = 1.24 $\AA$, $r_{\rm {Th^{4+}}}$ = 1.21 $\rm \AA$, and $r_{\rm {U^{4+}}}$ = 1.17 $\rm \AA$) \cite{Shannon}, and $r_{\rm Be^{\rm II}}$ is the Be metal-bonding radius (= 1.14 $\rm \AA$) \cite{Yang}, respectively. 
Here, we tentatively used the effective ionic radius for the 12-coordination-number site as $r_{\rm M}$ \cite{Shannon}, since there is no data for 24-coordination-number sites to our best knowledge. 
As seen in Fig. 6, $\theta_{\rm E}$ is robust to a change in $r_{\rm gfd}$ within the range of the error bar, suggesting that the above-mentioned interpretation for other cage-structured compounds is not simply applicable to the $M$Be$_{13}$ systems.

Another key parameter to discuss the low-energy phonon mode is the atomic mass. 
In the case of the Einstein oscillation, $k_{\rm B}$$\theta_{\rm E}$ can be rewritten to $\hbar$$\omega_{\rm E}$ $\propto$ ($k$/$m_{\rm_E}$)$^{1/2}$, where $\omega_{\rm E}$ is the Einstein frequency, $k$ is the spring constant, and $m_{\rm E}$ is the mass of the Einstein oscillator. 
Assuming that the $M$Be$_{13}$ compounds have similar values of $k$, it is expected that heavier mass of the $M$ atom (= $m_{\rm M}$) yields smaller $\theta_{\rm E}$. 
In this assumption, $\theta_{\rm E}$ for LaBe$_{13}$ ($m_{\rm {La}}$ $\sim$ 139) was estimated to be $\sim$ 210 K, when we calculate it from $\theta_{\rm E}$ of 160 K for UBe$_{13}$ ($m_{\rm U}$ $\sim$ 238). 
This estimated $\theta_{\rm E}$ for LaBe$_{13}$ deviates from the experimental value of $\sim$ 170 K. 
This disagreement may be attributed to the present simple assumption that $k$ has similar values among the $M$Be$_{13}$ compounds, since the value of $k$ generally depends on $r_{\rm gfd}$ and an ionic state of the guest ion. 
In this context, $\theta_{\rm E}$ in the $M$Be$_{13}$ systems might be determined by a combination of several parameters, such as $r_{\rm gfd}$, $m_{\rm M}$, and possibly electronic states, resulting in similar values of $\theta_{\rm E}$. 
Furthermore, in this discussion, the low-energy phonon mode was simply treated as the Einstein phonon, which is independent of the Debye phonon, although actual phonon dispersion should be more complicated. 
Recently, inelastic x-ray scattering measurements in SmBe$_{13}$ performed by Tsutsui $et$ $al$. revealed that a phonon dispersion curve of the Sm atom shows a flat part near a zone boundary. 
The energy of the flat part is approximately 16 meV, which is consistent with $\theta_{\rm E}$ of 157(10) K determined from the present XRD study. \cite{Tsutsui1}

Finally we comment on a relationship between the low-energy phonon mode and electronic states of conduction electrons in the $M$Be$_{13}$ systems. 
It has been theoretically proposed for cage-structured compounds that a possible enhancement of the quasiparticle mass is caused by the interaction between anharmonic vibration of the guest ion and conduction electrons. \cite{Hattori} 
The electronic specific-heat coefficient $\gamma$ for $M$Be$_{13}$ ($M$ = La, Sm, U and Th) are listed in Table II. 
The characteristic energy of the low-energy phonon mode appears to have no relation to the value of $\gamma$, implying that the HF state for UBe$_{13}$ may not originate from the presence of the low-energy phonon mode. 
However, it has also been suggested that an electron--phonon coupling between the conduction electrons and the low-energy Einstein phonon also plays an important role in formation of the phonon-mediated HF state. \cite{Yu, Kusunose, Tsutsui2} 
Therefore, it is necessary to check the strength of the electron--phonon coupling in the $M$Be$_{13}$ systems to reveal the role of the low-energy phonon mode in the HF state for UBe$_{13}$. 
To deepen our understanding of the low-energy phonon mode in the present systems, further studies, such as a systematic observation of the phonon dispersion and the strength of the electron--phonon coupling, are needed and now in progress.

\section{SUMMARY}
We measured XRD on powdered LaBe$_{13}$, SmBe$_{13}$, and UBe$_{13}$ at low temperatures in order to investigate their structure parameters and characteristics of the low-energy phonon. 
The obtained XRD patterns and the Rietveld refinements revealed that the present compounds keep a NaZn$_{13}$-type cubic structure with an almost ideal snub cube formed by 24 Be$^{\rm II}$ atoms involving the $M$ atom even at low temperatures down to 7 K. 
Furthermore, the present study provides crystallographic collateral evidence for the presence of the low-energy phonon modes common to the $M$Be$_{13}$ systems, which can be explained by an Einstein model. 
It is considered that the rigid Be$^{\rm II}$ framework can be treated as a Debye solid with $\theta_{\rm D}$ of 600--800 K, while the M ions behave like Einstein oscillators with $\theta_{\rm E}$ of $\sim$160 K. 
Interestingly, the obtained $\theta_{\rm E}$ values in the present systems, including the values reported thus far, appear to be independent of either $m_{\rm M}$ or $r_{\rm gfd}$ in the snub cube, which is a characteristic feature not found in other cage-structured compounds.

\begin{acknowledgments}
The authors thank Dr. S. Tsutsui for fruitful discussions. 
The present research was supported by JSPS KAKENHI Grants No. JP20224015(S), No. JP25400346(C), No. JP26400342(C), No. JP15H05882, and No. JP15H05885(J-Physics). 
\end{acknowledgments}

\end{document}